\documentclass[12pt]{article}
 \usepackage{latexsym, amsmath, epsfig, amsfonts, amsbsy}
 \DeclareFontFamily{OT1}{rsfs10}{}
 \DeclareFontShape{OT1}{rsfs10}{m}{n}{ <-> rsfs10 }{}
 \DeclareMathAlphabet{\mathscript}{OT1}{rsfs10}{m}{n}

 \else\oddsidemargin25mm\evensidemargin25mm\marginparwidth25mm\fi
 \def\Z{\mathbb{Z}}

 \def\bpl{\Big(}
 \def\bpr{\Big)}
 \def\e{\epsilon}
 
 \def\t{\theta}

 \def\der{\partial}
 \newcommand{\ft}[2]{{\textstyle\frac{#1}{#2}}}
 \def\brr{\begin{equation}}
 \def\err{\end{equation}}
 \def\brr{\begin{eqnarray}}
 \def\err{\end{eqnarray}}
 \def\ba{\left(\begin{array}}
 \def\ea{\end{array}\right)}
 \def\lf{\left.\begin{array}{c}}
 \def\rf{\end{array}\right.}

 \def\derbar{\stackrel{\leftrightarrow}{\partial}}
 \newcommand{\dr}{\raise.3ex\hbox{$\stackrel{\leftarrow}{\partial }$}{}}
 \newcommand{\dl}{\raise.3ex\hbox{$\stackrel{\rightarrow}{\partial}$}{}}
 \newcommand{\topi}{\raise.3ex\hbox{$\stackrel{\pi}{\longrightarrow}$}{}}
 \newcommand{\ns}{\normalsize}

 \renewcommand{\a}{\alpha}

 \renewcommand{\d}{\delta}
 \newcommand{\s}{\sigma}

\begin{document}

 \begin{titlepage}


\title{
   \hfill{\ns hep-th/0311095\\}
   \hfill{\ns HWS-2003A11\\}
   \hfill{\ns November 2003\\[2cm]}
   {\LARGE Duality and Central Charges\\[.1in]
   in Supersymmetric Quantum Mechanics\\[1cm]}}

\author{{\bf
   Michael Faux\footnote{faux@hws.edu}\,\, and
   Donald Spector\footnote{spector@hws.edu}}\\[5mm]
   {\it Department of Physics} \\
   {\it Hobart and William Smith Colleges} \\
   {\it Geneva, NY 14456} \\
   {\it USA}}

\date{}

\maketitle

\vspace{.3in}

 \begin{abstract}
 \noindent
 We identify a class of point-particle models that exhibit
 a target-space duality.  This duality arises from a construction
 based on supersymmetric quantum mechanics with a
 non-vanishing central charge. Motivated by analogies to string
 theory, we are led to speculate regarding mechanisms for
 restricting the background geometry. \\[.1in]
\end{abstract}

\thispagestyle{empty}

\end{titlepage}


 \noindent
 One phenomenon usually regarded as peculiarly ``stringy" is
 that of space-time duality.  The simplest instance
 occurs when a quantum string propagates in a ten-dimensional target space
 involving one compact circular spatial dimension having radius $R$.
 In this case, $T$-duality  \cite{alvarez} exchanges quantized
 Kaluza-Klein momentum modes with wrapping modes of the string,
 and at the same time replaces $R\to \a'/R$.
 Since wrapping modes are characteristic of extended objects,
 one might not expect to find a similar duality connecting
 ostensibly distinct {\it point} particle models with different
 classical background geometries.
 In this letter, we demonstrate that, surprisingly, a similar duality
 does hold for the case of a supersymmetric point particle propagating
 on a manifold with the topology of a cylinder,
 provided one incorporates a nontrivial central charge into the
 supersymmetry algebra. In this context,
 Kaluza-Klein modes are clearly present, and correspond to momenta of
 the particle directed around the cylinder.  But the wrapping modes of
 the string are replaced with a central charge parameter whose
 physical origin is at present obscure.

 Supersymmetric quantum mechanics \cite{witten} can be formulated as a one-dimensional
 supersymmetric quantum field theory; the lone dimension in this
 context is time.  A bosonic field is, then, a real-valued function of
 time, and a fermionic field is a Grassman-valued function of time.
 The $d=1$, $N=1$ superalgebra with a central charge
 is specified by the following relations,
 \brr \{\,Q\,,\,Q^\dagger\,\}=2\,H
      \hspace{.4in}
      [\,H\,,\,Q\,]=0
      \hspace{.4in}
      Q^2=Z \,,
 \label{susyalgebra}\err
 where $Q$ is the complex supercharge, $Z$ is the complex central
 charge, and $H$ is the Hamiltonian.  A consequence of the
 super-Jacobi identity is
 that $Z$ and $Z^\dagger$ each commute with $Q$ and
 with $Q^\dagger$.  Note that $[\,H\,,\,Q\,]=0$ must be
 specified as an independent condition when $Z\ne 0$.

 In higher dimensional field theories, central charges can
 appear in superalgebras owing to topological features of solitonic
 background field configurations \cite{wittenolive, hlouspec}.  In practice,
 these quantities arise as surface terms in integrals
 appearing in superalgebra anticommutators.  When formulating quantum
 mechanics as a field theory, however, there are no spatial integrals
 to produce such boundary terms.  Therefore,
 a topological explanation for the appearance of the central charge
 in (\ref{susyalgebra}) would require a suitably
 modified version of the usual field-theoretic explanation.
 One possibility in this regard would involve centrally-extended
 supersymmetric quantum mechanics as the natural description of
 effective physics localized on topological zero-branes present
 in higher dimensional field theories, such as a ``kink" soliton
 in a two-dimensional WZW model which has degenerate
 classical vacua \cite{graham1}, or a point-like intersection
 of two one-dimensional domain walls in a three-dimensional WZW
 model \cite{gt, pms}.

 When a charge is topological in origin, it is naturally quantized.
 For this reason, the scenarios described above conceivably
 provide a rationale for the quantization of the central charge term
 in (\ref{susyalgebra}).
 In this paper, however, we view this charge
 simply as an allowable extension to the algebra, and investigate
 the ramifications.  There is no {\it a priori} reason, from
 an algebraic perspective, that this charge should be quantized.
 Nevertheless, a duality we uncover in this paper
 suggests the existence of a more fundamental construction
 underlying the class of models we introduce, in which the
 central charge is quantized, perhaps
 in a manner similar to the examples described above.

 In this letter, we restrict attention to the case that
 the central charge $Z$ is real.  We do this for
 two reasons, which we believe may be related to each other.
 First, this restriction appears naturally when a two-dimensional
 (1,1) superalgebra is compactified to one
 dimension; in this case, the real central charge in the compactified
 algebra corresponds to the two-dimensional momentum component
 directed around the compactified dimension.
 Second, as we show in \cite{ShapeBPS},
 the above superalgebra with real central charge supplies
 a natural setting for understanding the shape
 invariance approach to exact solubility \cite{inhull, gendenshtein, coopgk, coopku}.

 In terms of ``pre-quantum" classical constructions, superalgebras are
 represented by transformation rules interrelating
 components of irreducible multiplets.  A
 supersymmetry transformation is expressed as
 $\d_Q(\e)=\e\,Q+\e^\dagger\,Q^\dagger$, where $\e$ is a complex
 Grassman parameter.  In terms of this operation, the $N=1$ superalgebra
 can be written, in the case of a real central charge, as
 \brr [\,\d_Q(\e_1)\,,\,\d_Q(\e_2)\,]=
      -4\,i\,\e_{[1}^\dagger\,\e_{2]}\,\der_t
      -2\,i\,(\,\e_1\,\e_2+\e_1^\dagger\,\e_2^\dagger\,)\,\d_Z \,,
 \label{algebra}\err
 where we have used the canonical representation of the Hamiltonian
 as the generator of time translations, $H=i\,\der_t$,
 and have defined a ``central charge transformation" $\d_Z$ via
 $Z=Z^\dagger=i\,\d_Z$.
 That irreducible representation which most economically
 includes a real central charge includes, off-shell, one real boson
 $T$, one complex fermion
 $\chi$, and one real auxiliary boson $B$. The
 supersymmetry transformation rules are given by
 \brr \d_Q\,T &=& i\,\e\,\chi+i\,\e^\dagger\,\chi^\dagger
      \nonumber\\[.1in]
      \d_Q\,\chi &=& \e^\dagger\,(\,\dot{T}+i\,B\,)
      +\mu\,\e
      \nonumber\\[.1in]
      \d_Q\,B &=& \e\,\dot{\chi}-\e^\dagger\,\dot{\chi}^\dagger
      \,,
 \label{basic}\err
 where a dot represents a time derivative.
 The inhomogeneous term, which appears in the transformation rule
 $\d_Q\chi$, includes a real parameter $\mu$,
 and is a feature novel to this paper, we believe.
 This term is associated with the central charge transformation
 $\d_Z\,T=\mu$.  The fields $\chi$ and $B$
 are invariant under the central charge.  Inspired by analogous
 constructions in higher-dimensional field theories,
 we refer to (\ref{basic}) as a
 ``vector" multiplet.

 An interesting model involves two
 vector multiplets, the first having a vanishing central charge
 and the second having a nonvanishing central charge
 implemented as in (\ref{basic}).  By denoting the components
 of the first of these multiplets using
 a subscript ``1"
 and the second using a subscript ``2",
 we therefore consider the following transformation rules,
 \brr \d_Q\,T_1 &=& i\,\e\,\chi_1
      +i\,\e^\dagger\,\chi_1^\dagger
      \hspace{.6in}
      \d_Q\,T_2 = i\,\e\,\chi_2
      +i\,\e^\dagger\,\chi_2^\dagger
      \nonumber\\[.1in]
      \d_Q\,\chi_1 &=&
      \e^\dagger\,(\,\dot{T}_1+i\,B_1\,)
      \hspace{.57in}
      \d_Q\,\chi_2 = \e^\dagger\,(\,\dot{T}_2+i\,B_2\,)
      +\mu\,\e
      \nonumber\\[.1in]
      \d_Q\,B_1 &=& \e\,\dot{\chi}_1
      -\e^\dagger\,\dot{\chi}_1^\dagger
      \hspace{.75in}
      \d_Q\,B_2 = \e\,\dot{\chi}_2
      -\e^\dagger\,\dot{\chi}_2^\dagger
      \,.
 \label{rules}\err
 In order to find an invariant action functional,
 we start with a supersymmetric sigma model, invariant when there
 is no central charge, and then append additional terms
 as needed to maintain supersymmetry invariance
 when the central charge parameter $\mu$ is non-zero.
 The action therefore is described by the superspace
 expression
 \brr S=\int dt\,d\t^\dagger\,d\t\,\bpl\,
      G_{ij}(V^1,V^2)\,D^\dagger V^i\,DV^j\,\bpr
      +\cdots \,.
 \err
 where $V^1$ and $V^2$ are superfields associated with the
 two vector multiplets, $\t$ is a complex Grassman superspace coordinate,
 $D={\der}/{\der\t}+i\,\theta^\dagger\,\der_t$
 is a superspace derivative,
 $G_{ij}(\,T_1\,,\,T_2\,)$ describes a metric on the target space,
 and the ellipsis represents terms, at higher-orders
 in $\mu$, needed to maintain supersymmetry in the presence of
 the central charge.
 The zeroth-order action, i.e. terms at order $\mu^0$,
 are straightforward to determine using standard superfield techniques.
 The correction terms, appearing at higher-orders in $\mu$, can be determined
 by careful analysis of the component action, using the
 transformation rules (\ref{rules}).

\begin{figure}
 \begin{center}
 \includegraphics[width=1.5in]{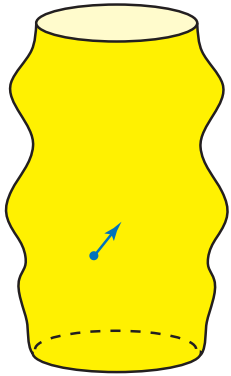}\\[.1in]
 \caption{A supersymmetric particle propagating on a rigid wiggly cylinder.
 The bosonic degrees of freedom describe the position of the particle
 on the cylinder, $T_1$ is the axial coordinate, and $T_2$ is
 the angular coordinate.  The fermionic degrees of freedom $\chi_i$ describe an
 internal ``state" of the particle, which one can interpret in terms
 of target-space spinor components.  The radius of the cylinder is
 $R(T_1)=h(T_1)^{1/2}$.}
 \label{wiggle}
 \end{center}
 \end{figure}

 For this paper, we specialize to the following
 class of Euclidean metrics,
 \brr ds^2=dT_1^2+h(T_1)\,dT_2^2 \,,
 \label{metric}\err
 where $h(T_1)$ is an arbitrary non-negative function.
 Thus, $G_{ij}={\rm diag}(\,1\,,\,h(T_1)\,)$.
 Furthermore, we take $T_2$ to be an angular
 coordinate, $T_2\in [\,0\,,\,2\,\pi\,]$ with endpoints
 identified.  In this case, the target space has the topology
 of a cylinder, having axial coordinate
 $T_1$ and a radius which depends on $T_1$ according to
 $h(T_1)^{1/2}$.  Thus, our model describes a
 supersymmetric particle propagating on a rigid wiggly cylinder,
 as shown in Figure \ref{wiggle}.
 In this case, the complete supersymmetric Lagrangian
 turns out to be
 \brr L &=&
      \ft12\,\dot{T}_1^2
      -\ft12\,i\,\chi_1^\dagger\derbar_t\chi_1
      +\ft12\,B_1^2
      \nonumber\\[.1in]
      & & +h(T_1)\,\bpl\,
      \ft12\,\dot{T}_2^2
      -\ft12\,i\,\chi_2^\dagger\derbar_t\chi_2
      +\ft12\,B_2^2\,\bpr
      \nonumber\\[.1in]
      & &
      +i\,h'(T_1)\,\chi_{[1}^\dagger\chi_{2]}\,\dot{T}_2
      -\ft12\,h''(T_1)\,\chi_2^\dagger\chi_2
      \chi_1^\dagger\chi_1
      \nonumber\\[.1in]
      & & +\ft12\,h'(T_1)\,\bpl\,
      \chi_1^\dagger\chi_2\,B_2
      +\chi_2^\dagger\chi_1\,B_2
      -\chi_2^\dagger\chi_2\,B_1\,\bpr
      \nonumber\\[.1in]
      & & -\ft12\,\mu\,i\,h'(T_1)\,
      (\,\chi_1\,\chi_2+\chi_1^\dagger\chi_2^\dagger\,)
      -\ft12\,\mu^2\,h(T_1) \,.
 \label{Lagrangian}\err
 The final line in (\ref{Lagrangian}) describes those
 special terms needed to maintain supersymmetry in the
 presence of the inhomogeneous term in the transformation rules
 (\ref{rules}).  At this point, the real parameter $\mu$
 enters the action as an arbitrary coupling strength; it has been
 introduced for the express purpose of inserting a central
 charge into the superalgebra.
 A model of our sort is specified
 by a choice of the parameter $\mu$ and by a choice of the background ``wiggle"
 function $h(T_1)$.

 The Dirac brackets derived from (\ref{Lagrangian}) provide the
 basic commutator and anti-commutator relationships
 which must be satisfied by the quantum operators
 corresponding to the canonical variables
 $T^i$, $P_i$, $\chi^i$ and $\chi^{i\,\dagger}$ .  In our case,
 the relevant expressions are
 \footnote{N.B. The indices on $T^i$ and $\chi^i$ are
 lowered by $\d_{ij}$, not by $G_{ij}$.}
 \brr [\,P_i\,,\,T^j\,] &=& -i\,\d_i\,^j
      \nonumber\\[.1in]
      \{\,\chi^i\,,\,\chi^{j\,\dagger}\,\} &=& G^{ij}
      \nonumber\\[.1in]
      [\,P_1\,,\,\chi^2\,] &=& \ft12\,i\,\frac{h'}{h}\,\chi^2 \,,
 \label{brackets}\err
 where $G^{ij}$ is the inverse of the target-space
 metric $G_{ij}$.
 All other basic (anti)commutators vanish.
 The first two lines in (\ref{brackets}) are generic, and the third is
 specific to the cylindrical geometry we have chosen in (\ref{metric}).
 Together, these determine $P_i=-i\,\der_i$ and
 \brr \chi^1 &=& \ft12\,(\Gamma^3+i\,\Gamma^4\,)
      \nonumber\\[.1in]
      \chi^2 &=& \ft12\,h^{-1/2}\,(\Gamma^1+i\,\Gamma^2\,) \,,
 \err
 where $\Gamma^\mu$ are Euclidean Dirac matrices, which satisfy the
 four-dimensional Clifford algebra
 $\{\,\Gamma^\mu\,,\,\Gamma^\nu\,\}=2\,\d^{\mu\nu}$.

 We determine the quantum operators corresponding to the
 conserved charges after first eliminating the auxiliary fields
 $B_1$ and $B_2$.  The supercharge operator, in this case, is
 given by
 \brr Q &=& P_1\,\chi_1
      +P_2\,\chi_2
      +\ft12\,i\,h':\chi_2\,\chi_2^\dagger\,\chi_1:
      +\mu\,h\,\chi_2^\dagger \,,
 \err
 and the central charge operator is given by
 \brr Z=\mu\,P_2 \,.
 \err
 These are the operator analogs of the classical supercharges
 determined from (\ref{Lagrangian}) using the Noether procedure.
 The ordering ambiguity in the fermion cubic term is resolved by
 imposing the superalgebra on the quantum operators.
 The quantum Hamiltonian is
 \brr H &=& \ft12\,P_1^2
      +\frac{1}{2\,h}\,P_2^2
      -i\,\frac{h'}{h}\,P_2\,\chi_{[1}^\dagger\chi_{2]}
      +\ft12\,:h\,{\cal R}\,\chi_1^\dagger\chi_1\chi_2^\dagger\chi_2:
      \nonumber\\[.1in]
      & & +\ft12\,i\,\mu\,h'\,
      (\,\chi_1\,\chi_2+\chi_1^\dagger\chi_2^\dagger\,)
      +\ft12\,\mu^2\,h \,,
 \label{hamiltonian}\err
 where ${\cal R}$ is the scalar curvature on the target space.
 In terms of the metric function $h(T_1)$, this is given by
 \brr {\cal R}=
      \ft12\,\bpl\,\frac{h'}{h}\,\bpr^2-\frac{h''}{h} \,.
 \err
 The term in the Hamiltonian (\ref{hamiltonian}) implicitly proportional to $h''$
 arises from the explicit
 fermion quartic in the Lagrangian (\ref{Lagrangian}),
 while the other implicit fermion quartic term arises after
 elimination of the auxiliary fields $B_i$.
 Accordingly, there are two independent ordering ambiguities in
 the fermion quartic terms in (\ref{hamiltonian}).
 These, too, are resolved by imposing the superalgebra
 (\ref{susyalgebra}) on the quantum operators.
 (This determines Weyl ordering on the fermion
 cubic term in $Q$ and determines the particular ordering on the
 fermion quartics in $H$ reflected in the expressions which
 follow.)

 It is useful to choose a particular representation for the
 fermions.  A convenient choice is given by
 \brr \Gamma^{1,2,3} = \ba{c|c}&\s_{1,2,3}\\\hline\s_{1,2,3}&\ea
      \hspace{.2in}
      \Gamma^4 = \ba{c|c}&-i\,{\bf 1}\\\hline i\,{\bf 1}&\ea \,,
 \label{basis}\err
 where $\s_{1,2,3}$ are the Pauli matrices.
 However, the properly-ordered Hamiltonian is not diagonal in this basis.
 A diagonal Hamiltonian is obtained by first computing the
 Hamiltonian in the basis (\ref{basis}), resolving the ordering
 ambiguities as described above, and then performing
 on all operators ${\cal O}$ the similarity transformation
 ${\cal O}\to\check{\cal O}=\Lambda^{-1}\,{\cal O}\,\Lambda$,
 where
 \brr \Lambda=\frac{1}{\sqrt{2}}\,\ba{cc|cc}
      1&0&-i&0\\i&0&-1&0\\\hline
      0&1&0&i\\0&i&0&1\ea \,.
 \err
 Now, since $h(T_1)^{1/2}$ is the radius of the cylinder,
 define $R(T_1)\equiv h(T_1)^{1/2}$.
 Also, since $T_2$ is an angular coordinate, it follows that
 the angular momentum $P_2$ is quantized,
 $P_2\equiv \nu\in\Z$.  In terms of these definitions,
 after some algebra, one can write the Hamiltonian
 associated with the sector having angular momentum
 $\nu$ as
 \brr \check{H}=\ba{cc|cc}A_+^\dagger\,A_++\mu\,\nu&&&\\
      &A_+\,A_+^\dagger+\mu\,\nu&&\\\hline
      &&A_-^\dagger\,A_-+\mu\,\nu&\\
      &&&A_-\,A_-^\dagger+\mu\,\nu \ea \,,
 \label{diagonal}\err
 where the operators $A_\pm$ are given by
 \brr A_\pm=\der_1+W'_\pm(T_1) \,,
 \err
 and the functions $W_\pm(T_1)$ are superpotentials
 induced by the background wiggle function $h(T_1)$
 and also by the central charge.  These are given by
 \brr W'_\pm(T_1) &=&-\frac{1}{2}\,\frac{R'}{R}
      \pm\,\bpl\,\frac{\nu}{R}-\mu\,R\,\bpr \,.
 \label{superpotentials}\err
 The four ``sectors" of the Hamiltonian (\ref{hamiltonian}),
 which we enumerate using an index
 $n$, each include a distinct scalar potential $V_n(T_1)$,
 determined from one of the superpotentials described by
 (\ref{superpotentials}).  There is an
 analogous four-sector Hamiltonian for each angular momentum sector,
 as labelled by the quantum number $\nu$.

 The Hamiltonian (\ref{diagonal}) has several features worthy of
 note. First, and foremost, the class of models we have introduced
 include a manifest target-space duality. Under the
 simultaneous transformations
 \brr R &\rightarrow& \frac{1}{R}
      \nonumber\\[.1in]
      \mu &\leftrightarrow& \nu
      \nonumber\\[.1in]
      T_1 &\to& -T_1 \,,
 \label{dualtran}\err
 all of the operators presented above transform simply.
 In particular, when $R(T_1)=R(-T_1)$, the transformation
 (\ref{dualtran}) leaves each of the
 operators $A_\pm$ multiplied by an overall minus sign, while
 for $R(T_1)=-R(-T_1)$, the operators $A_\pm$ map into each other.
 Thus, in these two cases, we have an explicit invariance of the
 Hamiltonian (\ref{diagonal}) under the mapping (\ref{dualtran}).

 Because the duality map (\ref{dualtran}) includes a swapping of
 $\nu$ and $\mu$, a more precise statement is that it connects
 the angular momentum sector $P_2=\nu$ in a model with central charge
 parameter $\mu$ with the angular momentum sector $P_2=\mu$
 in a model with central charge parameter $\nu$.
 If one interprets the quantum Hamiltonian (\ref{hamiltonian}) in terms
 of the sigma-model Lagrangian presented previously, $\nu$ is a quantized
 angular momentum, while $\mu$ is an arbitrary real parameter
 neither {\it a priori} quantized nor summed over.
 The duality relationship, as stated above, makes sense
 only if $\mu$ assumes integer values, however.
 This suggests to us the interesting possibility of an overarching theory,
 yet to be discovered, in which $\mu$ appears as an integer-valued
 topological quantity, properly summed over
 in the quantum theory. In such a construction,
 the exchange $\mu\leftrightarrow\nu$
 would shuffle different sectors of the same theory, and
 the mapping $R(T_1)\rightarrow 1/R(-T_1)$ would itself
 describe an invariance, since the theory would automatically
 contain separate sums over the $\mu$ sectors and the
 $\nu$ sectors.

 Looking at the superpotential functions (\ref{superpotentials}),
 one is also struck by the following observation:
 a global re-scaling of $R$ is equivalent to separate
 re-scalings of $\mu$ and $\nu$ which leaves the product
 $\mu\,\nu$ fixed. In other words,
 the Hamiltonian (\ref{hamiltonian}) has yet another
 invariance, as described by the operation
 \brr R &\to& \lambda\,R
      \nonumber\\[.1in]
      \nu &\to& \lambda\,\nu
      \nonumber\\[.1in]
      \mu &\to& \lambda^{-1}\,\mu \,,
 \err
 where $\lambda$ is an arbitrary real parameter.
 In this model, the mapping $\nu\rightarrow\nu+1$ already shifts
 among sectors, and if there is an overarching theory, there would
 be a sum over quantized $\mu$ values as well. These observations
 are consistent with the possibility that in the putative
 {\it\"uber}-theory, the quantities $\mu$ and $\nu$ appear as an
 electric and magnetic charge-pair, related to each other by an
 $SL(\,2\,,\,\Z\,)$ transformation analogous to a generalized
 electric-magnetic duality.

 We have exhibited a manifest ``T-duality" and have motivated
 a prospective ``S-duality" in a context we find
 elegant in its simplicity.  The appearance of a target-space duality
 within non-relativistic point-particle quantum
 mechanics is itself surprising.  But we also notice intriguing
 parallels with string theory.
 In the latter context, background geometry is famously restricted
 by quantum consistency conditions, conditions which are connected
 both to T-duality and to modular invariance.
 In the simpler models described in this paper,
 we also question what conditions might constrain the background
 geometry.  Restrictions based on the duality structures described
 above form an attractive basis for such speculation.
 Another possibility we find intriguing is based on
 shape invariance. As described at the beginning of this paper,
 centrally-extended super-algebras, which form the basis of our
 investigation,
 also comprise the natural context for shape invariant extensions
 to any solvable quantum mechanics \cite{ShapeBPS}.
 As it turns out, only particular background geometries
 can produce shape-invariant Hamiltonians of the form (\ref{hamiltonian}).
 Thus, we find that shape invariance, and the exact solubility to which it
 is associated, form an appealing mechanism for
 restricting the background geometry in which a supersymmetric
 particle can propagate.  In light of this, we are led to attempt
 to ponder the possibility of connections between shape-invariant
 world-line dynamics and an eventual elemental description of {\it M}-theory.

 \pagebreak

 \noindent
 {\bf Acknowledgements}\\[.1in]
 The authors are grateful to Ted Allen for careful
 tutelage into the vagaries of Dirac brackets and constrained
 dynamics, which proved most helpful in our analysis.
 M.F. is also grateful to M{\'a}ria and Emil Martinka,
 for hospitality, encouragement and halu{\v s}ky at the Slovak
 Institute for Basic Research, Podvazie Slovakia, where some
 of this work was performed.


\begin{thebibliography}{99}
 %
 \bibitem{alvarez}
 E. Alvarez, L. Alvarez-Gaum\'e and Y. Lozano,
 {\it An introduction to T duality in string theory},
 Nucl.\ Phys.\ Proc.\ Suppl.\ 41 (1995) 1.
 %
 \bibitem{witten}
 E. Witten,
 {\it Dynamical Breaking of Supersymmetry},
 Nucl.\ Phys.\ B188 (1981) 513-554.
 %
 \bibitem{wittenolive}
 E. Witten and D. Olive,
 {\it Supersymmetry Algebras that include Topological Charges},
 Phys.\ Lett.\ B78 (1978) 97.
 %
 \bibitem{hlouspec}
 Z. Hlousek and D. Spector,
 {\it Why Topological Charges Imply Extended Supersymmetry},
 Nucl.\ Phys.\ B370 (1992) 143-164 \,;\\
 Z. Hlousek and D. Spector,
 {\it Topological charges and
 central charges in 3+1 dimensional supersymmetry},
 Phys. Lett. B283 (1992) 75-79.
 %
 \bibitem{graham1}
 N. Graham and R.L. Jaffe,
 {\it Energy, Central Charge, and the BPS Bound
 for 1+1 Dimensional Supersymmetric Solitons},
 Nucl.\ Phys.\ B544 (1999) 432.
 %
 \bibitem{gt}
 G. Gibbons and P. Townsend,
 {\it A Bogmol'nyi Equation for Intersecting Domain Walls}
 Phys.\ Rev.\ Lett.\ 83 (1999) 1727.
 %
 \bibitem{pms}
 P.M. Saffin,
 {\it Tiling With Almost BPS Junctions}
 Phys.\ Rev.\ Lett.\  83 (1999) 4249.
 %
 \bibitem{ShapeBPS}
 M.Faux and D.Spector,
 {\it An Algebraic Basis for
 Energy Bounds in Shape-Invariant
 Supersymmetric Quantum Mechanics},
 HWS-2003B11.
 %
 \bibitem{inhull}
 L. Infeld and T.D. Hull,
 {\it The Factorization Method},
 Rev.\ Mod.\ Phys 23 (1951) 21-68.
 %
 \bibitem{gendenshtein}
 L. Gendenshtein,
 {\it Derivation of Exact Spectra of the Schr{\"o}dinger Equation
 by Means of Supersymmetry},
 JETP Lett. 38 (1983) 356-359.
 %
 \bibitem{coopgk}
 F. Cooper, J.N. Ginocchio and A. Khare,
 {\it Relationship between supersymmetry and solvable potentials},
 Phys.\ Rev.\ D36 (1986) 2485-2473.
 %
 \bibitem{coopku}
 F. Cooper, A. Khare, and U. Sukhatme,
 {\it Supersymmetry and Quantum Mechanics},
 Phys.\ Rept.\ 251 (1995) 267.
 %
 \end{thebibliography}
 \end{document}